\documentclass[aps,prb,twocolumn,showpacs,groupedaddress]{revtex4}
\usepackage{graphicx}
\usepackage{amssymb}
\usepackage{dcolumn}

\begin{document}

\title{Structural Anomalies at the Magnetic and Ferroelectric Transitions in $RMn_2O_5$ ($R$=$Tb$, $Dy$, $Ho$)}
\author{C. R. dela Cruz$^1$, F. Yen$^1$, B. Lorenz$^{1}$, M. M. Gospodinov$^{2}$, \\C. W. Chu$^{1,3,4}$,
W. Ratcliff$^5$, J. W. Lynn$^5$, S. Park$^6$, and S-W. Cheong$^6$}
\affiliation{$^{1}$Department of Physics and TCSUH, University of
Houston, Houston, TX 77204-5002} \affiliation{$^{2}$Institute of
Solid State Physics, Bulgarian Academy of Sciences, 1784 Sofia,
Bulgaria} \affiliation{$^{3}$Lawrence Berkeley National Laboratory,
1 Cyclotron Road, Berkeley, CA 94720} \affiliation{$^{4}$Hong Kong
University of Science and Technology, Hong Kong, China}
\affiliation{$^5$NIST Center for Neutron Research, NIST,
Gaithersburg, Maryland 20899} \affiliation{$^6$Department of Physics
$\&$ Astronomy and Rutgers Center for Emergent Materials, Rutgers
University, Piscataway, NJ 08854}
\date{\today }

\begin{abstract}
Strong anomalies of the thermal expansion coefficients at the
magnetic and ferroelectric transitions have been detected in
multiferroic $RMn_2O_5$. Their correlation with anomalies of the
specific heat and the dielectric constant is discussed. The results
provide evidence for the magnetic origin of the ferroelectricity
mediated by strong spin-lattice coupling in the compounds. Neutron
scattering data for $HoMn_2O_5$ indicate a spin reorientation at the
two low-temperature phase transitions.
\end{abstract}

\pacs{65.40.De,75.25.+z,75.80.+q,77.80.-e,77.84.Bw} \maketitle










Frustrated magnetic systems have recently attracted the attention of
solid state physicists with regard to the observed magneto-electric
couplings and the induced ferroelectricity in some orthorhombic rare
earth manganites
\cite{kimura:05,hur:04,hur:04b,kagomiya:03,kobayashi:05} and other
frustrated compounds.\cite{lawes:05} These materials are of
fundamental interest since it has been demonstrated that an external
magnetic field can rotate the ferroelectric (FE)
polarization.\cite{kimura:03,hur:04b,lawes:05} Despite intense
experimental investigations, the physical origin of the large
magneto-electric coupling and the ferroelectricity arising at the
magnetic lock-in transitions is not yet understood. The giant
magneto-dielectric effects and the ferroelectricity in these
compounds require the existence of sizable atomic displacements and
structural distortions, the magnetic origin of which is believed to
lie in extraordinarily strong spin-lattice interactions. The
experimental verification of these structural anomalies is essential
to prove the suggested intimate correlation between the magnetic and
FE orders.

The search for structural distortions by neutron scattering
experiments have indicated some anomalies in the temperature
dependence of the lattice parameters of $TbMn_2O_5$ \cite{chapon:04}
and $DyMn_2O_5$ \cite{blake:05}, however, due to the limited
resolution of such scattering experiments a unique assignment to the
various phase transitions in these compounds appears extremely
difficult. Other investigations of $HoMn_2O_5$ \cite{blake:05} and
$YMn_2O_5$\cite{kagomiya:03,kobayashi:04} have failed completely to
resolve structural anomalies. It is, therefore, one of the key
issues to detect and to characterize the structural distortions in
$RMn_2O_5$ manganites giving rise to ferroelectricity and to
investigate the coupling between magnetic, dielectric, and lattice
degrees of freedom.

In this communication we present our results of high-resolution
thermal expansion measurements along the three crystallographic
orientations in $RMn_2O_5$ ($R$=$Ho$, $Dy$, $Tb$). Sharp structural
anomalies are detected at all magnetic and FE phase transitions, the
strongest anomalies appearing at the transitions into the FE phases
associated with a significant change of the electric polarization.
Single crystals of $RMn_2O_5$ have been prepared by the
high-temperature solution growth method as described
elsewhere.\cite{mihailova:05,hur:04} Thermal expansion measurements
were conducted employing a high-resolution capacitance dilatometer.
The experimental techniques have been described
earlier.\cite{johansen:86} With our current device we achieved a
resolution (above the noise level) of 4 {\AA}. The expansion data
were correlated to anomalies of the $b$-axis dielectric constant and
the heat capacity measured for the same crystals.

In orthorhombic $RMn_2O_5$ the spins of the $Mn^{4+}$ and $Mn^{3+}$
ions and the $R^{3+}$ moments are coupled via the predominantly
antiferromagnetic (AFM) superexchange interactions giving rise to a
complex magnetic phase diagram.\cite{blake:05} The common features
for all $RMn_2O_5$ are a transition into a high-temperature N\'{e}el
phase (HTIC) with a two-component incommensurate (IC) magnetic
modulation characterized by a wave vector
$\overrightarrow{q}=(q_{x},0,q_{z})$ at $T_{N1}\approx$ 43 K
followed by a lock-in transition into a commensurate (CM) phase with
$\overrightarrow{q}=(0.5,0,0.25)$ at $T_{C1}$ only a few degrees
lower where ferroelectricity arises. With further decreasing
temperature, at $T_{C2}$, the CM phase becomes unstable towards a
low-temperature IC phase (LTIC) with a magnetic modulation
$\overrightarrow{q}\approx(0.48,0,0.3)$.
\cite{blake:05,chapon:04,wilkinson:81,kobayashi:04,kobayashi:04b,kobayashi:05}
The transition at $T_{C2}$ is accompanied by a significant decrease
of the FE polarization and is often referred to as a second FE phase
transition. In some rare earth compounds (e.g. $HoMn_2O_5$,
$DyMn_2O_5$) additional anomalies have been observed indicating an
even more complex magnetic structure that has yet to be
explored.\cite{ratcliff:05,mihailova:05}

It is characteristic for all $RMn_2O_5$ compounds that the various
magnetic phase changes are reflected in sharp and distinct anomalies
of the dielectric constant, as shown in Figs. 1a to 3a for $R$=$Ho$,
$Dy$, and $Tb$. This is a clear indication of strong
magneto-electric coupling due to large spin-lattice interactions.
The thermodynamic signature of the various transitions is obvious
from the peaks of the specific heat $C_{p}$, Figs. 1a to 3a. Several
transitions show pronounced hysteresis effects, not shown in the
figures (for enhanced clarity only warming data are included in
Figs. 1 to 4). The thermal expansivities, displayed in Figs. 1b to
3b, exhibit distinct anomalies at all magnetic transitions with the
strongest peaks observed at the FE transitions. Our data provide
striking evidence for the existence of extraordinarily large
spin-lattice interactions in $RMn_2O_5$ resulting in macroscopic
displacements along all three crystallographic axes. They further
prove that the magnetic and lattice degrees of freedom are strongly
coupled and the simultaneous magnetic and FE transitions at $T_{C1}$
and $T_{C2}$ have to be considered as the phase change of one highly
correlated system.

The thermal expansion data for the three manganites ($R$=$Ho$, $Dy$,
$Tb$) exhibit similarities but also distinct differences depending
on the rare earth ion. The changes of the lattice parameters have to
be discussed with respect to the magnetic orders derived from
neutron scattering
experiments.\cite{blake:05,ratcliff:05,wilkinson:81,chapon:04,kobayashi:04b}
The first transition from the paramagnetic (PM) and paraelectric
(PE) phase into the HTIC phase at $T_{N1}\approx$ 43 K is
characterized by a peak of $C_{p}$, an increase of $\varepsilon$,
and a sudden increase of the expansivities along $a$- and $c$-axes.
Considering the spiral IC magnetic order below $T_{N1}$ with the
wave vector $\overrightarrow{q}=(0.5+\beta,0,0.25+\delta)$, the
increase of $\alpha_{a}$ and $\alpha_{c}$ with the onset of AFM
order can be explained by the magnetoelastic effect. There is no
anomaly of $\alpha_{b}$ which is consistent with the lack of the
magnetic modulation along $b$.

\begingroup
\squeezetable
\begin{table}
\caption{Relative change of lattice parameters at the FE transitions
in $RMn_2O_5$. $\Delta L=L(T<T_C)-L(T>T_C)$.}
\begin{ruledtabular}
\begin{tabular}{l|cc|cc|cc}
$\Delta L/L$ & $HoMn_2O_5$ & & $DyMn_2O_5$ & & $TbMn_2O_5$ \\
($\times 10^{-6}$)& $T_{C1}$ & $T_{C2}$ & $T_{C1}$ & $T_{C3}$ & $T_{C1}$ & $T_{C2}$ \\
\hline
$\Delta a/a$  & 2.1   & 13.4  & 3.8   & -67.1  & -1.1  & 11.3 \\
$\Delta b/b$  & 2.5   & 5.7   & 2.0   & 16.3   & -2.0  & 11.5 \\
$\Delta c/c$  & -6.6  & -8.3  & -2.0  & 63.0   & 10.2  & -41.5 \\
\end{tabular}
\end{ruledtabular}
\end{table}
\endgroup

The FE transition at $T_{C1}$ coincides with the lock-in of the
magnetic wave vector into commensurate values. It is well marked by
sharp peaks of $C_p$ and $\varepsilon$ as well as strong peaks of
the expansivities along all three axes (Fig. 1 to 3). We define the
abrupt length change, $\Delta L$, at the transition as the
difference of the lengths in the low- and high-temperature phases
extrapolated to $T\gtrsim T_C$ and $T\lesssim T_C$, respectively.
The measured values for $\Delta L$ along $a$, $b$, and $c$ are
listed in table 1. For $R$=$Ho$ and $Dy$ the $c$-axis contracts
while $a$ and $b$ expand at $T_{C1}$ (Figs. 1b and 2b). However, the
opposite is observed for $R$=$T$b (Fig. 3b). This behavior has its
origin in the different spin modulations along $c$, given by
$q_z=0.25+\delta$. For Ho and Dy $\delta$ is negative
\cite{blake:05,ratcliff:05} but for Tb $\delta$ is
positive.\cite{chapon:04,kobayashi:04b} The length change of the
$c$-axis is directly correlated with the deviation of the magnetic
wave length $\lambda_m$ along $c$ from its commensurate value of
4$c$. The spin-lattice coupling leads to an attraction between the
IC spin wave and the underlying lattice period along $c$ between
$T_{N1}$ and $T_{C1}$ resulting in a stress on the lattice. This
stress is tensile for $R$=$Ho$, $Dy$ ($\lambda_m$$>$4$c$) but
compressive for $R$=$Tb$ ($\lambda_m$$<$4$c$). At the lock-in
transition the stress is suddenly released ($\lambda_m$=4$c$)
resulting in an abrupt decrease (increase) of $c$ for $R$=$Ho$, $Dy$
($Tb$). This transition at $T_{C1}$ is of first order as indicated
by the volume change derived from the expansion data, the jump of
$q_x$ and $q_z$,\cite{kobayashi:04b} and the thermal hysteresis
observed in $\varepsilon(T)$. Mediated by the elastic forces of the
lattice, $a$ and $b$ contract ($Tb$) or expand ($Ho$, $Dy$) opposite
to $c$. In $Ho$- and $TbMn_2O_5$ the length change along $c$
dominates whereas in $DyMn_2O_5$ it is rather minor (table 1). This
qualitative difference has its origin in the $Mn^{4+}$ spin
alignment along $c$ that polarizes the rare earth moments. For
$R$=$Ho$, $Tb$ there are alternating ferromagnetic (FM) and AFM
links along $c$ that may cause a positional modulation of $R$ and
the connecting oxygen, O(2), which is seen in $\alpha _c$ at
$T_{C1}$ whereas in $DyMn_2O_5$ the $Mn^{4+}$ spins are only FM
along $c$ and the expansion anomaly is accordingly
small.\cite{blake:05} Strong lattice anomalies have also been
observed in $GdMnO_3$ at the transition into the FE phase that
happens in this multiferroic compound at higher magnetic
fields.\cite{Baier:06}

The superexchange interactions between neighboring $Mn^{3+}$ and
$Mn^{4+}$ ions in the $a$-$b$ plane are all AFM and the smallest
closed loop of neighboring $Mn$ spins involves 5 ions. Therefore,
magnetic frustration must exist as, for example, revealed by the
spin order in $HoMn_2O_5$ displayed in the inset of Fig.
4.\cite{blake:05} The spins of two $Mn^{3+}$ per unit cell are each
frustrated with two neighboring $Mn^{4+}$ with the same spin
direction. Reducing this frustration by moving the two $Mn^{3+}$
away from the $Mn^{4+}$ (along the arrows labeled \textbf{P} in Fig.
4) generates a dipolar moment $P$ between the $Mn^{3+}$ and the
surrounding oxygen ions. The $a$-components of $P$ cancel out while
the $b$-components add up to the macroscopic polarization along $b$
and ferroelectricity. The proposed displacement lowers the symmetry
to $Pb2_1m$. This empirical picture is supported by recent
M\"{o}ssbauer experiments on $YMn_2O_5$ showing that neighboring
$Mn^{3+}$ ions are magnetically inequivalent.\cite{kagomiya:05} The
corresponding displacement vector of only two of the four $Mn^{3+}$
is a superposition of the basis vectors of the $\Gamma_{1g}$ and
$\Gamma_{3u}$ irreducible representations of the space group $Pbam$
in contrast to the model proposed for $YMn_2O_5$ where the
displacement of all four $Mn^{3+}$ ions was assumed ($\Gamma_{3u}$
representation).\cite{kagomiya:03} We would like to emphasize that
the AFM modulation along $a$ with $q_x$=0.5 is crucial for the above
discussion, as it leads to the frustration and displacement of both
$Mn^{3+}$ and the net polarization along $b$. Considering the role
of magnetic frustration to stabilize ferroelectricity in $RMn_2O_5$
there are interesting similarities to multiferroic $Ni_3V_2O_8$ and
$TbMnO_3$. For the latter compounds it was shown that the transition
from sinusoidal to helical magnetic modulation can introduce a third
order coupling giving rise to FE order.\cite{lawes:05,
kenzelmann:05} By symmetry the same argument holds also for two
non-collinear spin density waves with the same propagation vector
but different phases.\cite{mostovoy:06} In fact, neutron scattering
experiments have revealed the existence of a non-collinear spin
structure in $RMn_2O_5$ ($R$=$Ho$, $Dy$, $Tb$)\cite{blake:05} with
$\overrightarrow{q}$ along the $a$-axis which shows that
ferroelectricity below $T_{C1}$ is not forbidden by symmetry.
However, it is not clear yet if the magnetic structure between
$T_{C1}$ and $T_{N1}$ is sinusoidal and the transition into the FE
phase follows the same mechanisms as in $Ni_3V_2O_8$ or $TbMnO_3$.
Furthermore, the magnetic modulation in the FE phase of $RMn_2O_5$
is commensurate whereas it is incommensurate in $Ni_3V_2O_8$ or
$TbMnO_3$.

The FE transition at $T_{C1}$ is followed by additional phase
changes at lower $T$ as indicated by characteristic changes of
$\varepsilon$ and $C_p$ and distinct anomalies of the expansion
coefficients (Figs. 1 to 3). For $R$=$Ho$ and $Dy$ a step-like
increase of both, $\varepsilon$ and $C_p$, at $T_{N2}$=22 K and 27
K, respectively, is followed by another increase of $\varepsilon$ at
$T_{C2}$=16 K (Ho) and 17 K (Dy) with a significant change of the FE
polarization.\cite{hur:04} The $\varepsilon$-anomaly at $T_{C2}$ is
largest in $HoMn_2O_5$ but there is no equivalent signature in $C_p$
of $Ho(Dy)Mn_2O_5$. Recent neutron scattering experiments on
$DyMn_2O_5$ indicate that a spin reorientation takes place at
$T_{N2}$ and also at $T_{C2}$.\cite{ratcliff:05} Our neutron
scattering results shown in Fig. 4 provide clear evidence for spin
reorientations in $HoMn_2O_5$ at $T_{N2}$ and $T_{C2}$. The
integrated intensity of the (1.5,0,0.25) peak suddenly increases at
both transitions whereas that of the (0.5,0,2.25) peak shows a sharp
drop. The details of the magnetic orders in the various phases have
yet to be determined. In $TbMn_2O_5$ the spin reorientation
transition at $T_{N2}$ appears to be missing, however, from the
distinct step of $C_p$ at 24 K (similar to the $C_p$ anomalies in
$Ho$- and $DyMn_2O_5$ at $T_{N2}$) we conclude that this transition
coincides with the FE transition at $T_{C2}$ (determined by the
sharp increase of $\varepsilon$).

The FE transition at $T_{C2}$ is accompanied by another change of
the magnetic modulation that becomes incommensurate again,
$\overrightarrow{q}\approx(0.48,0,0.3)$ (LTIC phase). The strongest
anomalies of the expansion coefficients appear at $T_{C2}$ for
$R$=$Ho$ and $Tb$ (Figs. 1 and 3, table 1) indicating a sizable
distortion of the lattice due to the CM $\rightarrow$ LTIC phase
change. The corresponding anomalies are less pronounced in
$DyMn_2O_5$ (Fig. 2). Instead, another low-$T$ transition at
$T_{C3}$=7.5 K is associated with giant changes of the lattice
constants, a large contraction of the $a$-axis and an expansion of
$b$ and $c$ (table 1). This phase transition is the onset of strong
AFM order of the $Dy$-moments along $a$
($\overrightarrow{q}_{(Dy)}$=(0.5,0,0)) \cite{ratcliff:05} and the
$a$-contraction is explained by large magnetostrictive effects.
$C_p$ exhibits a sharp peak at $T_{C3}$ and $\varepsilon$ drops to
its high-temperature value ($T$$>$$T_{N1}$). It was suggested that
the low-$T$ state in $DyMn_2O_5$ is paraelectric
\cite{higashiyama:04} and it appears conceivable that the lattice
strain below $T_{C3}$ is dominated by the Dy order, as evidenced by
the huge expansion anomalies at $T_{C3}$. The AFM order of the Dy
moments with $q_x$=0.5 appears to dominate the low-temperature
magnetic structure and the associated lattice distortions which
could explain the PE low-$T$ phase in $DyMn_2O_5$. However, neutron
scattering experiments have found that the Dy magnetic order
coexists with the CM and IC orders of the $Mn$
spins\cite{ratcliff:05} and further investigations are needed for a
final conclusion.

The strongest lattice anomalies observed in our thermal expansion
measurements occur at the FE transitions involving a significant
change of the polarization. They are highly anisotropic and prove
the existence of an extraordinarily strong spin-lattice coupling.
The magnetic and FE orders have to be considered as one highly
correlated system that is coupled by magnetoelastic interactions.


\begin{acknowledgments}
This work is supported in part by NSF Grant No. DMR-9804325, the
T.L.L. Temple Foundation, the J. J. and R. Moores Endowment, and the
State of Texas through the TCSUH and at LBNL by the DoE. The work of
M. M. G. is supported by the Bulgarian Science Fund, grant No F
1207. The work at Rutgers is supported by NSF-DMR-0080008.
\end{acknowledgments}

\begin{figure}
\caption{(Color Online) Anomalies at the magnetic phase transitions
in $HoMn_2O_5$. The transition temperatures are marked by vertical
dotted lines. Only warming data are shown in all figures. (a)
Dielectric constant along $b$ (open circles, left scale) and
specific heat (closed circles, right scale). (b) Thermal
expansivities along $a$-, $b$-, and $c$-axes. For better clarity the
data along $a$- and $b$-axis are offset by 2.6 and 1.4 units,
respectively.}
%
\caption{(Color Online) Same as Fig. 1 for $DyMn_2O_5$. The $b$-axis
expansivity is offset by 0.6 units. Note the different scales (left
and right) chosen for the expansivities below and above 10 K,
respectively.}
%
\caption{(Color Online) Same as Fig. 1 for $TbMn_2O_5$. The $a$- and
$b$-axis expansivities are offset by 5 and 4.2 units, respectively.}
%
\caption{(Color Online) Integrated intensities of two neutron
scattering peaks of $HoMn_2O_5$ along different directions in
reciprocal space. Inset: $Mn$-spin configuration in the CM phase
below $T_{C1}$.}
\end{figure}





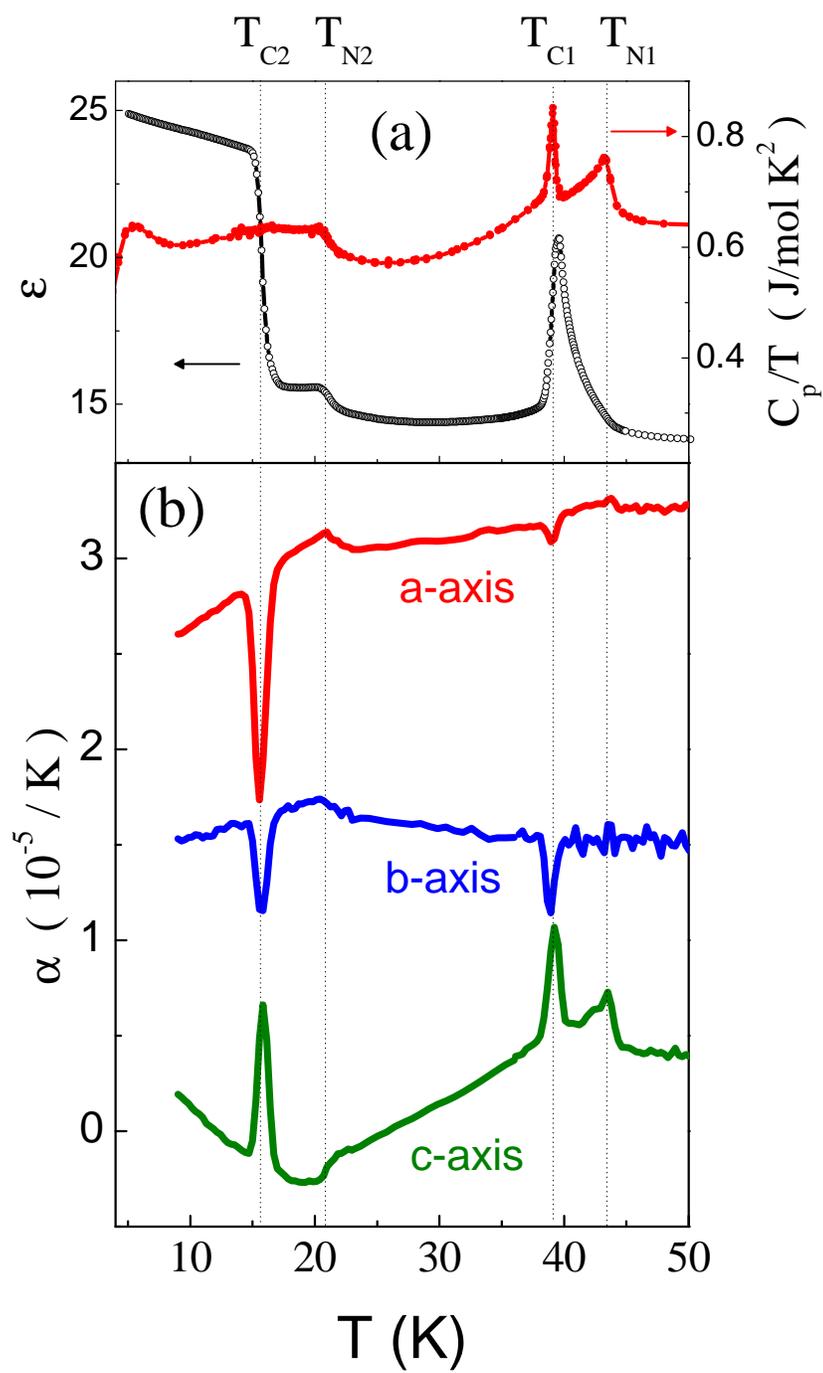
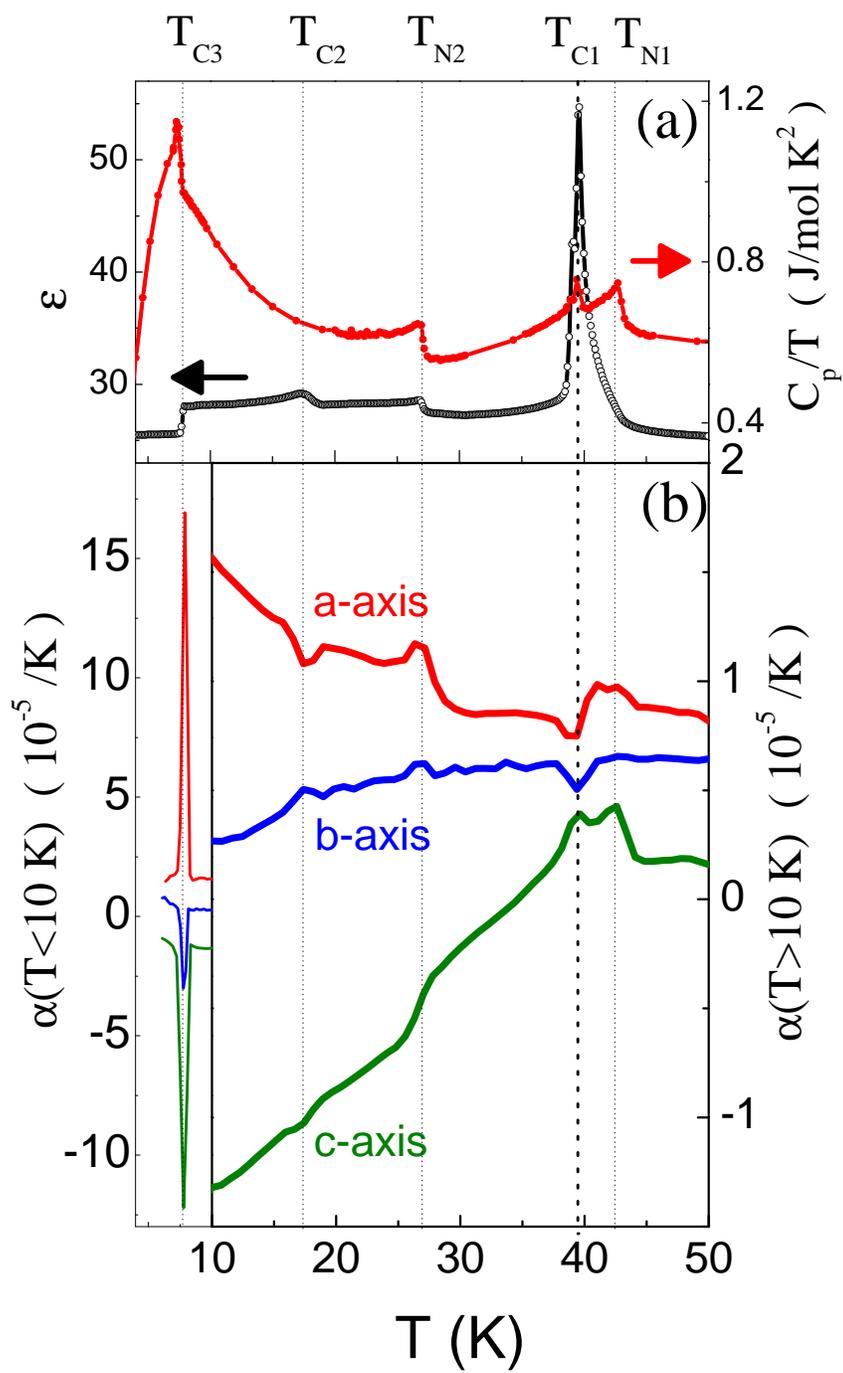
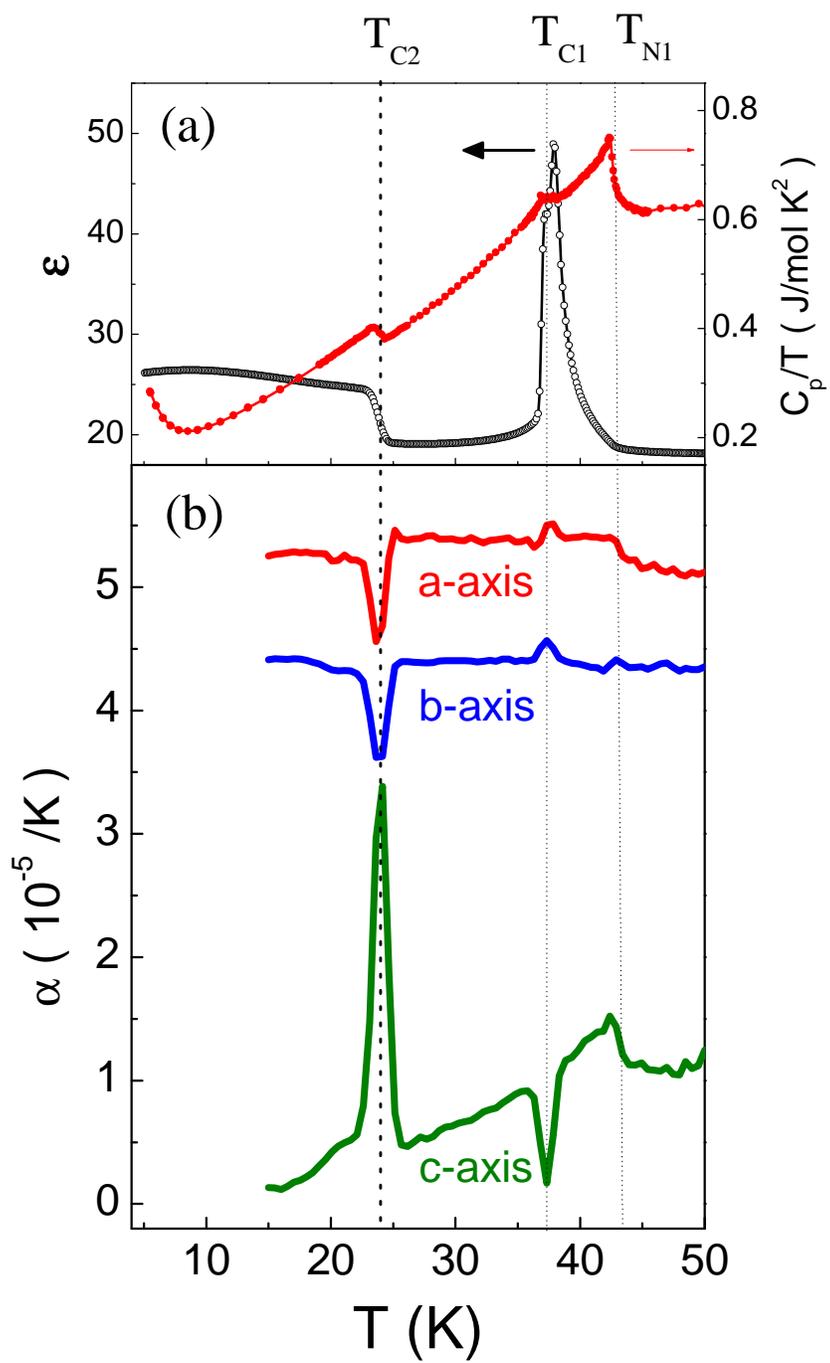
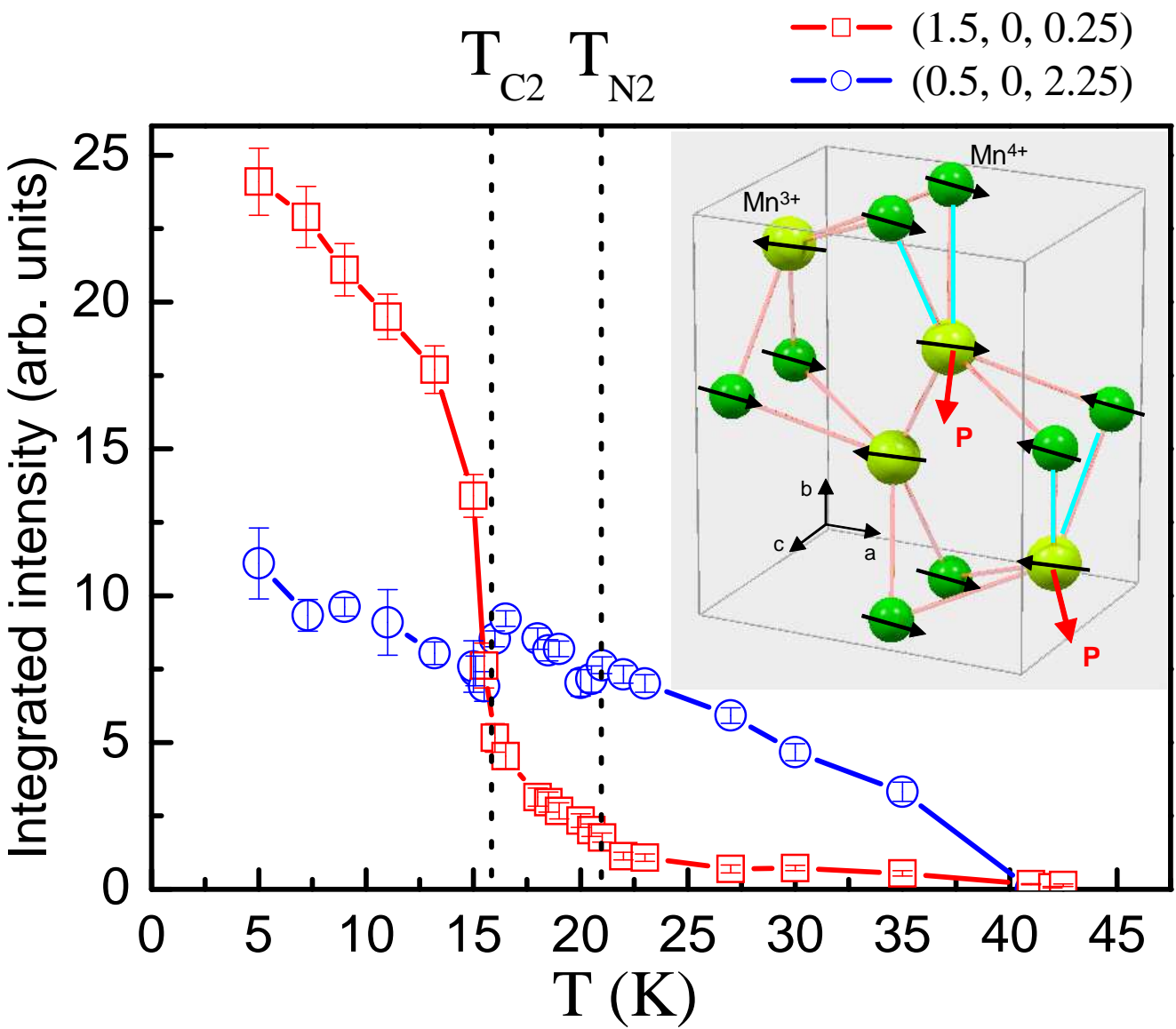
\end{document}